\documentclass[11pt]{article}
\usepackage{times}
\usepackage{sidecap}
\usepackage{geometry}
\usepackage{floatrow}
\usepackage{float}
\usepackage{hyperref}
\usepackage{graphics,graphicx}
\usepackage{subcaption}
\usepackage{epsf,wrapfig}
\usepackage{amsmath}
\usepackage{upgreek}

\usepackage{natbib,bm}

\usepackage{paralist}
\let\olditem\item
\renewenvironment{thebibliography}[1]{%
  \section*{\refname}
  \let\par\relax
  \renewcommand{\item}[1][]{\olditem[\textbullet]}%
  \inparaenum}{\endinparaenum}

\newcounter{aff}

\DeclareGraphicsExtensions{.jpg,.pdf,.png,.eps}

\usepackage{caption}
\usepackage{xcolor}

\usepackage{tcolorbox}

\bibpunct{(}{)}{;}{a}{}{,}

\usepackage{soul}

\definecolor{darkcyan}{rgb}{0.0, 0.0, 0.7}

\definecolor{boxcolor}{RGB}{193, 229, 255}
\definecolor{figboxcolor}{RGB}{193, 229, 255}
\definecolor{boxcolorgray}{RGB}{227, 227, 227}
\definecolor{darkcyan}{rgb}{0.0, 0.0, 0.7}

\sethlcolor{boxcolorgray}

\def\gtrsim{\lower 2pt \hbox{$\, \buildrel {\scriptstyle >}\over
{\scriptstyle \sim}\,$}}
\def\lesssim{\lower 2pt \hbox{$\, \buildrel {\scriptstyle <}\over
{\scriptstyle \sim}\,$}}

\def\approxlt{\lower.2em\hbox{$\buildrel < \over \sim$}}
\def\approxgt{\lower.2em\hbox{$\buildrel > \over \sim$}}

\def\be{\begin{equation}}
\def\ee{\end{equation}}
\def\bea{\begin{eqnarray*}}
\def\eea{\end{eqnarray*}}

\def\jref@jnl#1{{\rm#1}}
\def\aj{\jref@jnl{AJ}}			
\def\araa{\jref@jnl{ARA\&A}}		
\def\apj{\jref@jnl{ApJ}}			
\def\apjl{\jref@jnl{ApJ}}		
\def\apjs{\jref@jnl{ApJS}}		
\def\ao{\jref@jnl{Appl.~Opt.}}		
\def\apss{\jref@jnl{Ap\&SS}}		
\def\aap{\jref@jnl{A\&A}}		
\def\aapr{\jref@jnl{A\&A~Rev.}}		
\def\aaps{\jref@jnl{A\&AS}}		
\def\azh{\jref@jnl{AZh}}			
\def\baas{\jref@jnl{BAAS}}		
\def\jrasc{\jref@jnl{JRASC}}		
\def\memras{\jref@jnl{MmRAS}}		
\def\mnras{\jref@jnl{MNRAS}}		
\def\pra{\jref@jnl{Phys.~Rev.~A}}	
\def\prb{\jref@jnl{Phys.~Rev.~B}}	
\def\prc{\jref@jnl{Phys.~Rev.~C}}	
\def\prd{\jref@jnl{Phys.~Rev.~D}}	
\def\pre{\jref@jnl{Phys.~Rev.~E}}	
\def\prl{\jref@jnl{Phys.~Rev.~Lett.}}	
\def\pasp{\jref@jnl{PASP}}		
\def\pasj{\jref@jnl{PASJ}}		
\def\qjras{\jref@jnl{QJRAS}}		
\def\skytel{\jref@jnl{S\&T}}		
\def\solphys{\jref@jnl{Sol.~Phys.}}	
\def\sovast{\jref@jnl{Soviet~Ast.}}	
\def\ssr{\jref@jnl{Space~Sci.~Rev.}}	
\def\zap{\jref@jnl{ZAp}}			
\def\nat{\jref@jnl{Nature}}		
\def\iaucirc{\jref@jnl{IAU~Circ.}}
\def\aplett{\jref@jnl{Astrophys.~Lett.}}
\def\apspr{\jref@jnl{Astrophys.~Space~Phys.~Res.}}
\def\bain{\jref@jnl{Bull.~Astron.~Inst.~Netherlands}}
\def\fcp{\jref@jnl{Fund.~Cosmic~Phys.}}
\def\gca{\jref@jnl{Geochim.~Cosmochim.~Acta}}
\def\grl{\jref@jnl{Geophys.~Res.~Lett.}}
\def\jcp{\jref@jnl{J.~Chem.~Phys.}}	
\def\jgr{\jref@jnl{J.~Geophys.~Res.}}	
\def\jqsrt{\jref@jnl{J.~Quant.~Spec.~Radiat.~Transf.}}
\def\memsai{\jref@jnl{Mem.~Soc.~Astron.~Italiana}}
\def\nphysa{\jref@jnl{Nucl.~Phys.~A}}
\def\physrep{\jref@jnl{Phys.~Rep.}}
\def\physscr{\jref@jnl{Phys.~Scr}}
\def\planss{\jref@jnl{Planet.~Space~Sci.}}
\def\procspie{\jref@jnl{Proc.~SPIE}}

\def\ion#1#2{#1$\;${\small\rm\@Roman{#2}}\relax}

\newcount\lecurrentfam
\def\LaTeX{\lecurrentfam=\the\fam \leavevmode L\raise.42ex
\hbox{$\fam\lecurrentfam\scriptstyle\kern-.3em A$}\kern-.15em\TeX}
\def\sizrpt{
(\fontname\the\font): em=\the\fontdimen6\font, ex=\the\fontdimen5\font
\typeout{
(\fontname\the\font): em=\the\fontdimen6\font, ex=\the\fontdimen5\font
}}
\definecolor{dark}{rgb}{ 0.34, 0.09, 0.49}

\usepackage{fancyhdr,url}
\usepackage{transparent}
\usepackage{tablefootnote}
\usepackage{mathpazo}

\newcommand{\cfa}{Center for Astrophysics $|$ Harvard \& Smithsonian, 60 Garden Street, Cambridge, MA 02138, USA}
\newcommand{\bhi}{Black Hole Initiative, Harvard University, 20 Garden Street, Cambridge, MA 02138, USA}

\begin{document}

\noindent
\begin{raggedright}
{\Huge\bfseries High-frequency water megamasers as precision probes of black holes, galaxies and cosmology \\[0.5em]
\Large A case for a next-generation sub-mm/mm interferometer
}\end{raggedright}\medskip

\noindent
\begin{flushleft}
Ingyin Zaw$^{1*}$, C.M. Violette Impellizzeri$^{2*}$, Dominic W. Pesce$^{3,4}$, Elizabeth Humphreys$^5$, Hannah~R.~Stacey$^5$, Sophie Venselaar$^6$, Chengyu Kuo$^7$, Jack Gallimore$^8$, Yoshiaki Hagiwara$^9$, Satoki Matsushita$^{10}$, Jim Braatz$^{11}$
\end{flushleft}

\vspace{0.5em}
{
\footnotesize{
\noindent $^*$ Co-lead authors \\
\noindent $^1$ New York University Abu Dhabi, P.O. Box 129188, Abu Dhabi, United Arab Emirates, \\
$^2$ ASTRON, Netherlands Institute for Radio Astronomy, Oude
Hoogeveensedijk 4, Dwingeloo, 7991 PD, The Netherlands, \\
$^3$ \cfa, \\
$^4$ \bhi, \\
$^5$ European Southern Observatory (ESO), Karl-Schwarzschild-Str. 2, D-85748 Garching bei München, Germany,\\
$^6$ University of Geneva, Department of Astronomy, ch. d’Ecogia 16, 1290 Versoix, Switzerland, \\
$^7$ Physics Department, National Sun Yat-Sen University, No. 70, Lien-Hai Rd, Kaosiung City 80424, Taiwan, R.O.C, \\
$^8$ Department of Physics and Astronomy, Bucknell University, Lewisburg, PA 17837, USA, \\
$^9$ Natural Science Laboratory, Toyo University, 5-28-20, Hakusan, Bunkyo-ku, Tokyo 112-8606 Japan, \\
$^{10}$ Institute of Astronomy and Astrophysics, Academia Sinica, 11F of Astronomy-Mathematics Building, AS/NTU, No.1, Sec.4, Roosevelt Rd., Taipei 106319, Taiwan, R.O.C.,\\
$^{11}$ National Radio Astronomy Observatory, 520 Edgemont Rd., Charlottesville, VA 22903, USA } \\
}
\bigskip
\begin{tcolorbox}[boxsep=3pt,left=0pt,right=0pt,top=1pt,bottom=1pt,colframe=boxcolor,colback=boxcolor]

\justify
\textbf{Abstract:} 
Water megamasers in the circumnuclear disks of active galactic nuclei are among the most precise probes of supermassive black holes. VLBI imaging of 22\,GHz disk masers has provided geometric distances to independently constrain $H_0$, accurate black hole mass measurements, and sub-parsec views of accretion disk geometry. ALMA has opened a new window into water maser transitions in the sub-mm/mm regime, which trace temperatures, densities and disk regions that are inaccessible at 22\,GHz, and can be observed at higher spatial resolution and with less free-free absorption. However, current detections are limited to a few bright extragalactic masers in nearby galaxies. We present the case for a next-generation sub-mm/mm interferometer - ALMA2040 - that would open a new parameter space for maser studies. We show that a 10-fold increase in line sensitivity and three times longer baselines compared to ALMA would transform sub-mm maser studies from a few bright detections into a population-level probe, delivering black hole masses and geometric distances from sub-mm masers alone, independent of and complementary to 22\,GHz measurements. Integrated into a VLBI array, ALMA2040 would resolve maser disks out to $>$160\,Mpc, extend maser black hole masses into the largely untested intermediate-mass regime, reveal magnetic field structure in accretion disks through polarization studies, and test systematics in $H_0$ measurements.
\end{tcolorbox}

\newpage{}

\vspace{-0.7cm}
\section{Scientific context and open questions}
Water megamaser emission is a powerful probe of warm, dense gas in the central parsec of active galactic nuclei (AGN, see \cite{Lo2005} for a review). Individual clouds of Earth-sized maser spots can be resolved in position and velocity with Very Long Baseline Interferometry (VLBI) in Keplerian rotation about supermassive black holes (disk masers), in jets and outflows (jet masers), or, occasionally, both. Disk masers have been used to make precise and accurate measurements of {\it geometric} distances, supermassive black hole (SMBH) masses, and accretion disk geometries in galaxies within $\sim$150 Mpc. Such measurements have made significant contributions to cosmology, galaxy evolution, and AGN physics. While most studies so far have been carried out at 22\,GHz, sub-mm water megamaser studies, coupled with a next-generation sub-mm/mm interferometer (hereafter, ALMA2040), would complement, if not rival, the scientific constraints from current and future cm-wave observatories.



\subsection{Geometric distances and $\mathbf{H_0}$}

{The Hubble Constant, $H_0$, the rate of the Universe's current expansion, was first estimated nearly a century ago by Edwin Hubble, and its value has been refined and hotly debated ever since. Currently, the measurements from observations of the early Universe, the cosmic microwave background (CMB), and baryon acoustic oscillations (BAO) with big bang nucleosynthesis (BBN), are in sharp tension with those from the late Universe, differing by $5\sigma$ to $7\sigma$ (\citealt{CAI2026} and references therein). These discrepancies between early and late times may reflect new fundamental physics. Many solutions have been proposed (\citealt{DiValentino:2021}, for review). However, whether the observed discrepancies truly reflect new physics or, rather, measurement errors in one or more of the methods requires further theoretical and observational work. 

Most late-Universe measurements rely on ``standard candles'', such as Cepheids and Type~Ia supernovae (e.g., \citealt{Freedman2001, Riess2022, Casertano2026}). Rubin/LSST will enable constraints on $H_0$ to $<1$\% precision from standard candles \citep{Jones:2019}. However, standard candles rely on assumptions and may be vulnerable to unknown systematics. Masers, instead, can be used to measure geometric distances by tracking the motion of gas in Keplerian rotation around a black hole (BH; \citealt{Pesce_2020a}), directly in the Hubble flow. Currently, the value based on six megamaser galaxies of $H_0 = 73.9 \pm 3.0$ km\,s$^{-1}$ Mpc$^{-1}$ \citep{Pesce2020} agrees with other late-Universe measurements, but its uncertainty is still too large to exclude the early-Universe value. Since the systematic errors of $H_0$ from each maser system are uncorrelated, the precision on $H_0$ based on $N$ systems decreases as $\sqrt(N)$. The typical error for a single galaxy is $\sim$5-10\%, implying a sample of $\sim$25-100 for a 1\% $H_0$ measurement. Improvements will come not only from additional systems, but also from better precision in the disk modeling of currently known systems. If the distance error can be reduced to $\sim$3\%, only $\sim$10 systems would be needed to reach a precision of 1\% on $H_0$.


\subsection{Maser masses: BH-galaxy co-evolution and intermediate mass black holes (IMBHs)}

Two fundamental constituents in galaxy evolution are i) the correlation between SMBH masses and host galaxies at low masses and ii) the abundance of intermediate-mass black holes (IMBHs) in the Universe. Currently, better understanding of both of these fields is stymied by the lack of precise and accurate mass measurements in low-mass galaxies. 

It is well established that all massive galaxies, with masses comparable to or larger than the Milky Way, host SMBHs at their center and that the SMBH masses scale with the properties of their host galaxies such as stellar velocity dispersion ($\sigma$), bulge mass ($M_{Bulge}$) and stellar mass ($M_*$) (e.g., \citealt{Kormendy2013}). The AGN fraction in massive galaxies is also well studied and can reach tens of percent, depending on the selection criteria (e.g., \citealt{Kauffmann2003, Hickox2018ARA&A}. However, in low-mass galaxies, the situation is less clear. 

Theoretical models make different predictions for the BH occupation fraction at low masses, ranging from 50\% to 100\% (e.g. \cite{Gallo2023}). Very few nuclear BHs in low-mass galaxies are found to be accreting (e.g. \cite{Bellovary2019}), and predictions of BH-galaxy scaling relations at the low-mass end depend on the models of BH seed formation and growth in the early Universe (e.g. \cite{Volonteri2010, Volonteri2021}). 

Although intermediate mass black holes (IMBHs; $M_{BH} \sim 10^2$--$10^5\,M_\odot$) which bridge the mass gap between stellar-mass BHs ($M_{BH} \sim 10$--$100\,M_\odot$) and SMBHs ($M_{BH} \gtrsim 10^5\,M_\odot$), are predicted to exist to seed the growth of SMBHs \citep{Volonteri2021} their abundance is highly debated and, again, depends on BH seed formation and growth (e.g., \citealt{Mezcua2017, Volonteri2021}). If they exist, some should be at the centers of low-mass galaxies. BHs in low-mass galaxies are generally smaller, which leads to smaller spheres of influence and lower Eddington luminosity, so they are harder to detect both dynamically and through signatures of continuous accretion. There are only a handful of BHs in low-mass galaxies with dynamical mass measurements, and most other methods, such as single-epoch optical spectra, have errors of at least a factor of a few \citep{Greene2020}. 


Disk megamasers provide the most precise measurements of IMBH masses, but only one low-mass AGN, IC~750, is known to host one. Its mass, $1.4 \times 10^5$ M$_\odot$, falls at the edge of the IMBH range and lies two orders of magnitude below the $M_{BH}$--$\sigma$ relation and one to two orders of magnitude below the $M_{BH}$--$M_{\rm bulge}$ and $M_{BH}$--$M_\star$ relations \citep{Zaw2020}. Such systems are difficult to detect and resolve: maser detection rates fall with galaxy mass \citep{Rosenthal2020}, and smaller BHs will have more compact maser disks. Major gains in sensitivity  and resolution would be needed to detect additional systems in order to understand whether this object is an outlier or signals a change to our picture of BH formation and growth.

\subsection{Accretion disk geometry and gas dynamics}
\label{maser_disk}

Water megamasers also provide a unique means of probing the physical conditions, dynamics, and structure of molecular gas in AGN accretion disks. Those scales are prohibitive to access at high resolution, especially for thermal molecular gas emission, which is typically observed with ALMA.  Megamasers, in contrast, display exceptionally high brightness temperatures, allowing observation at milliarcsecond (mas) resolution and thus sub-pc scales. Recent high-sensitivity VLBI observations of NGC~1068 at 22\,GHz, for example, have detected three times more masers compared to original VLBI imaging \citep{Greenhill1996}, demonstrating the impact of improved sensitivity in maser studies. The masers revealed a complex distribution, tracing a nearly edge-on disk with Keplerian velocity around a ($\sim1.7\times10^7\,M_\odot$) BH \citep{Gallimore2023}. These observations additionally revealed for the first time the existence of so-called `spokes', i.e. narrow, nearly parallel maser filaments, suggestive of an ordered magnetic field threading the accretion disk. Even more direct evidence for the role of magnetic fields was obtained by the first direct detection of linearly polarized water masers \citep{Gallimore2024}. Precise \(0.3\,\mathrm{mas}\) astrometry on the continuum was key to registering the exact location of the SMBH with its mid-infrared counterpart (3–13 $\mu$m; \citealt{Violeta2022}), and thus the dust distribution seen by MATISSE and GRAVITY \citep{Violeta2022, Gravity2020}. Finally, recently emerging evidence shows that masers may be lifted from the disk mid-plane, in a way that is reminiscent of magnetically supported "prominences" in the Sun. These findings clearly highlight the need for improved line sensitivity and for more lines of sight through the disk, requiring additional sub-mm lines. 


 \begin{figure}
\vspace{-10pt}
\includegraphics[width=\linewidth]{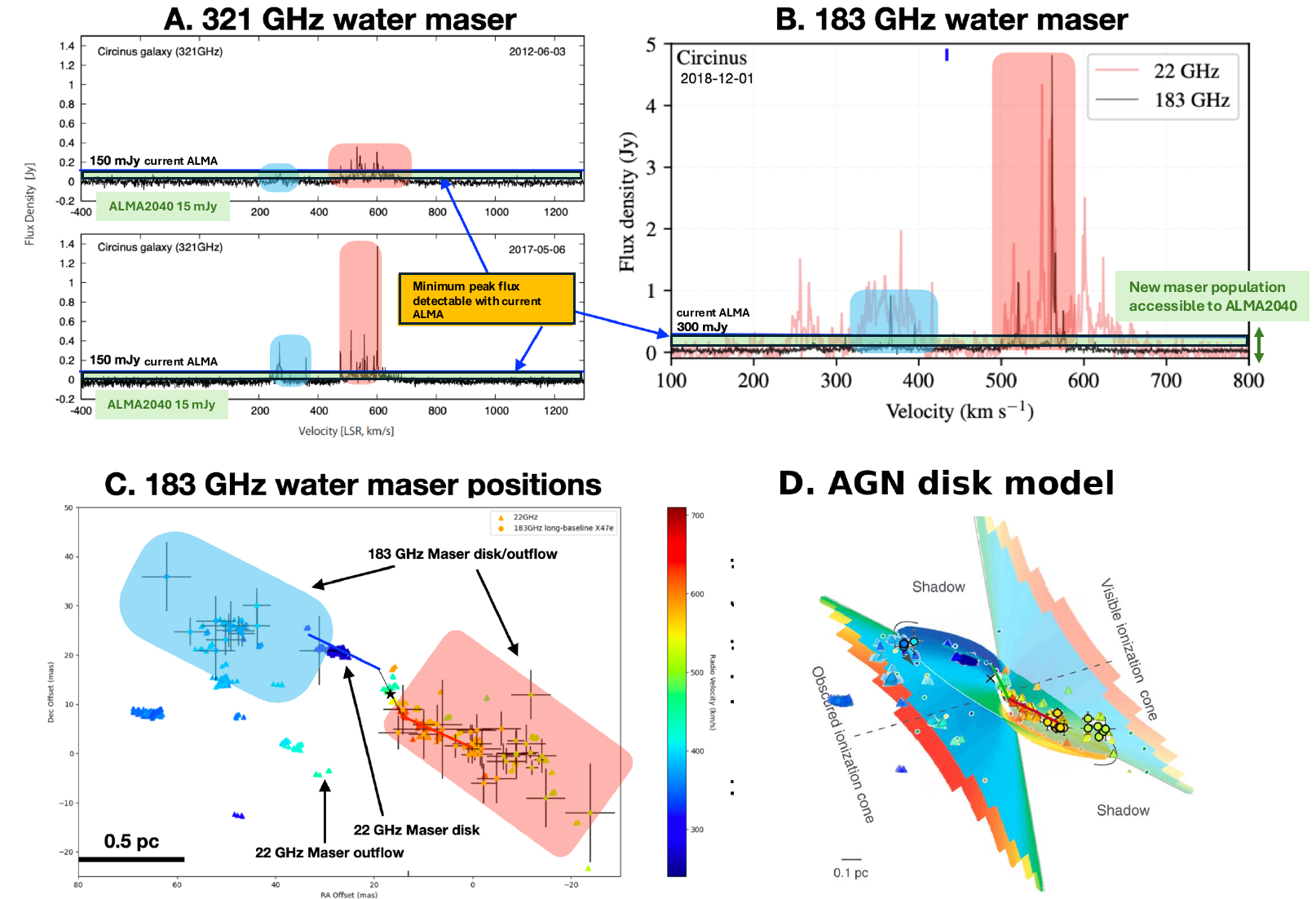}
      \caption{Sub-mm masers observed with current ALMA capabilities. ALMA2040 will shift from detecting the brightest masers to mapping the full disk accretion flow. (A–B) Observed 321 and 183 GHz H\(_2\)O maser spectra; (C) Resolved spatial distribution of the 183\,GHz masers compared to 22\,GHz disk and outflow; (D) Schematic disk/outflow geometry (Venselaar et al. in prep).}
    \label{fig:1} 
\end{figure}
 
\section{Sub-mm/mm maser emission}
\label{sec:submm} 

Until very recently, extragalactic megamaser science was dominated by the 22\,GHz ($6_{16}-5_{23}$) transition of water. This was historically due to the availability of cm-wavelength receivers for VLBI, combined with the difficulty of observing at sub-mm wavelengths, where atmospheric water vapour requires an exceptionally dry, high-altitude site. However, it has long been known that the water molecule has a plethora of additional maser transitions, which remain largely unexplored in extragalactic sources. These span both low-energy states (e.g., the 183\,GHz line at $E_u \sim 200$\,K) and relatively high-energy states (e.g., the 321\,GHz line at $E_u \sim 1860$\,K).  Additional transitions of note include those with rest frequencies 325, 437, 439, 471 and 658\,GHz, which have been observed to be strong Galactic masers \citep{Gray2016}. 

Access to high frequencies is critical for the 658\,GHz transition, which arises from the first excited bending-mode state ($v_2=1$) of water, $\sim$2300--2400\,K. Maser models predict strong inversion of this line in hot, dense gas, with kinetic temperatures $\sim$1000--1500\,K and $n({\rm H_2O})\sim10^5$--$10^6$\,cm$^{-3}$ \citep{Gray2016}. The 658\,GHz line is expected to trace the hottest and potentially innermost molecular gas able to survive close to the BH, at twice the resolution of the 321\,GHz line ($\sim$2\,mas for 50\,km baselines). At the other extreme, lower-excitation transitions, such as the 183\,GHz line, trace cooler material at larger radii and heights than the 22\,GHz masers, as already demonstrated in Circinus (Sect.~\ref{maser_disk}).  Broad frequency coverage, extending to Band~1, is additionally essential for high-redshift science (Sect.~\ref{sec:highz}). Together with predicted masers e.g., at 899 and 922\,GHz, these lines span the full ALMA frequency range.

Some of the high-frequency water megamasers have been detected towards multiple extragalactic targets, e.g., those at 183, 321 and 380\,GHz \citep[e.g.,][]{Pesce2016, Pesce2023, Hagiwara2013, Hagiwara2021, Yang2023}.  ALMA has made first studies of sub-mm water maser lines possible, but so far only in a handful of nearby galaxies.  A 10-fold increase in line sensitivity will be required to make the science systematic. 




\subsection{From individual maser spots to complete accretion disks}
\label{maser_disk}

High-frequency masers are crucial probes of accretion disks, not only because of the potential for higher angular resolution compared to 22\,GHz, but also because different transitions require different temperature and density conditions \citep{Gray2016}, which will arise from different regions of the circumnuclear disk. 

Recent observations of the Circinus galaxy have enabled, for the first time, to resolve the sub-mm maser disk (see Figure~\ref{fig:1}C from Venselaar et al. in prep.). Furthermore, multiple sub-mm line detections (183, 321 and 325\,GHz; Fig.~\ref{fig:1}A,B), could all be traced to the circumnuclear disk. The resolved 183\,GHz emission has a larger vertical extent and occurs at larger radii than the 22\,GHz masers, implying cooler, less dense material; 321\,GHz, on the other hand, traces hotter, more compact regions, potentially mapping the innermost molecular disk and thus marking the transition between the parsec-scales and the accretion flow feeding the BH. 

The idea of ``super masers'' is also an interesting one \citep{Gray2016} as it rests on models of water maser ``backbone'' lines (22, 183, 380\,GHz). They predict that, depending on saturation, the 380\,GHz line could outshine the 22\,GHz line (an earlier similar prediction comes from \citealt{Neufeld1991}). There is work ongoing to find new super masers with ALMA, but a major limiting factor is line sensitivity. Propagating the level of detail achieved for nearby galaxies to a larger sample of AGN will require enabling sub-mm maser searches at comparable sensitivity achieved in the cm-regime. 



\subsection{Maser polarization}
\label{sec:polarization}

The magnetic field properties near the accretion disk and outflows provide insights into jet launching and BH feeding, and these can be traced by masers. Because water is a non-paramagnetic molecule, circular (Zeeman) polarization of maser emission is expected to be weak ($<1\%$), but linear polarization $>1\%$ can arise from anisotropic pumping and/or saturation and directly traces the plane-of-sky field orientation. Modeling of the maser emission \citep{Lankhaar2019} predicts that sub-mm maser lines become strongly polarized once they saturate, with both the 321\,GHz and 183\,GHz maser transitions becoming significantly polarized ($\gtrsim 3\%$) once saturated, with the 183\,GHz line reaching even higher fractional polarization (up to $\sim$14\%). Moreover, the sub-mm masers largely avoid the Faraday depolarization that has hampered polarization work at 22\,GHz, making them a cleaner probe of the magnetic field close to the BH \citep{Surcis2020}. Additionally, recent polarization modeling predicts strong anisotropic pumping for the 183\,GHz transition capable of generating non-Zeeman circular polarization \citep{Lankhaar:2024}.


Maser polarization, however, is also intrinsically sensitivity limited. Measurements of extragalactic maser emission are effectively restricted to the brightest components ($\gtrsim150$\,mJy), yielding only a handful of polarization vectors (Figure~\ref{fig:1}A,B). Recovering the coherent magnetic-field geometry across the disk, i.e., the magnetic-field tomography of these sub-parsec environments, will test whether the field follows the rotating disk, traces vertical or radial structures, or changes at the disk–wind interface.

\subsection{Towards high-z maser emission} \label{sec:highz}

The same backbone transitions providing motivation for sub-mm searches offer a route to extending the megamaser technique to cosmological distances. The 380\,GHz line can be shifted into current ALMA Bands 1-3-6-7, which will allow VLBI observations to resolve high-z maser disks at $\sim 20–40\,\upmu$as. These cosmological masers can provide an independent geometric, distance-ladder-independent anchor, as well as accurate BH masses deep into the early Universe. The 22\,GHz line alone is not enough; being intrinsically faint, it is easily swamped by the stronger radio continuum, as was the case for the first (and to date only) megamaser detection in the lensed quasar MG~0414+0534 at $z=2.6$ \citep{Impellizzeri2008,Castangia2011}. 

The 380\,GHz line has the potential to outshine the 22\,GHz line \citep{Maloney2002,Gray2016}. However, this line emission from nearby galaxies cannot be observed from the ground due to atmospheric opacity; it can only be observed when redshifted. Indeed, \citet{Tarchi2024} reported the first unambiguous detection of 380\,GHz megamaser emission in TXS2226-184, revealing striking similarities with the 22\,GHz line. Evidence is also emerging at higher redshifts: a recent systematic ALMA survey of 12 lensed HyLIRGs at $z=1.8$–3.5 detected this line in 9/12 targets (83\%), with narrow spectral components pointing to genuine maser emission (Kuo et al. in prep.).


Current high-z 380\,GHz programs are nevertheless sensitivity-limited: reaching the $\approx$0.4\,mJy rms needed for a $5\sigma$ gigamaser detection already costs tens of hours per target, restricting us to the most extreme, most massive Compton-thick quasars. An order-of-magnitude sensitivity gain would open up the broader high-z AGN population to statistical studies.

\section{ALMA2040: Technical requirements}

\subsection{Requirements for key science}

A major increase in both line sensitivity and angular resolution relative to current ALMA is essential to make resolved measurements of maser disk kinematics on sub-pc scales, BH masses, and ultimately geometric distances. These requirements are coupled: the astrometric precision of an unresolved maser scales approximately as $\sigma_{\rm pos}\propto {\rm FWHM_{beam}}/{\rm SNR}$, so a 10$\times$ gain in sensitivity combined with a 3$\times$ finer beam at fixed flux density improves the positional precision by a factor of $\sim$30. For example, at 321\,GHz a maser currently detected at ${\rm SNR}=10$ in a $\sim$15\,mas beam has $\sigma_{\rm pos}\approx0.9$\,mas, whereas the same maser with ALMA2040 (${\rm SNR}=100$, $\sim$5\,mas beam) achieves $\sigma_{\rm pos}\approx0.03$\,mas ($30\,\upmu$as).

ALMA2040 will be able to reach these goals at larger distances. ALMA2040 alone would be able to resolve and detect a maser disk to $\sim$40\,Mpc ($1$\,mas is $\simeq0.1$\,pc at 20\,Mpc), with VLBI baselines extending to even larger distances, thanks to improved resolution. This is particularly important for intermediate-mass BHs, for which the sphere of influence and Keplerian maser disk scale with BH mass ($r\propto M_{\rm BH}/v^2$). So for example, the $\sim$0.2\,pc ($\sim$3\,mas) disk of IC\,750 is smaller than the $\sim$4--6\,mas ALMA2040 beam at 321\,GHz but can be mapped by centroiding velocity-resolved components at ${\rm SNR}\gtrsim10$ ($\sigma_{\rm pos}\sim0.3$\,mas).

For ALMA2040, the sensitivity gain is required not only to detect additional maser galaxies but also to recover the faint maser complexes necessary to constrain disk geometry and dynamics (Section~\ref{maser_disk}). Maintaining a high spectral resolution is also critical because maser lines are typically extremely narrow. Spectral resolutions of at least $\sim$0.1--0.5\,km\,s$^{-1}$ are required together with broad instantaneous bandwidth to encompass the systemic and high-velocity maser complexes.

Full-Stokes capability with excellent polarization purity is required to take advantage of the few-percent linear polarization expected from sub-mm water masers. Detecting 3\% polarization at $3\upsigma$ requires ${\rm SNR}\gtrsim100$ in Stokes~I, so the same $\sim10\times$ improvement in line sensitivity will enable polarization measurements of sources down to $\sim$15\,mJy (compared with $\sim$150\,mJy today), opening magnetic-field tomography across the disk. Instrumental polarization purity must be well below the expected signal ($\lesssim$0.3\%). 

Ultra-wide frequency coverage is a key requirement. The water molecule hosts a rich ladder of potentially masing transitions spanning different excitation temperatures and conditions, providing complementary probes of the temperature, density, and geometries. At $z\sim0$, the 96, 183, 321/325/354, 437--471, and 621/658\,GHz lines fall in ALMA Bands~3, 5, 7, 8, and~9, and the 899/922\,GHz lines in Band~10. Redshift moves these lines to lower bands; the 380\,GHz line is accessible in Band~7 for $z\gtrsim0.02$ and reaches Band~3 at $z\sim2.3$--3.5, and 183\,GHz reaches Band~1 at $z\gtrsim2.7$ (Sect.~\ref{sec:highz}). ALMA2040 should therefore retain access to Bands~1--10, with Bands~5--9 of highest priority for nearby disks (see Section~\ref{maser_disk}).

Integration of ALMA2040 into global VLBI arrays is an essential requirement. The much smaller beam provided by VLBI is needed to resolve maser disks on sub-parsec scales beyond the local Universe, and thus be able to estimate distances. ALMA2040 must improve on the current ALMA capabilities to operate as a phased array. The inclusion of a small number of larger-diameter antennas within ALMA2040 could be considered a powerful route to increasing instantaneous phasing efficiency. High-frequency VLBI requires deriving phase solutions within the short coherence time ($\leq$10\,s in Band~7, shorter than the $\approx$18\,s phasing loop on current ALMA; \citealt{Crew2023}). Larger collecting elements provide higher instantaneous sensitivity and enable phasing on fainter targets, which is particularly important for the frequency phase transfer (FPT) experiments, where lower-frequency continuum emission (e.g., at $\sim$86\,GHz) is tracked simultaneously with high-frequency emission, extending the coherence time. In the simulations presented below (See Fig.\ref{fig:2}, FPT from compact continuum emission at $\sim$86\,GHz extends the coherence time and allows disks to be resolved at distances where conventional high-frequency VLBI fails (e.g., at 160\,Mpc). ALMA2040 should, therefore, provide simultaneous ultra-wide-bandwidth observing.

\subsection{Simulated data}
\label{simulations}

\begin{figure}[t]
    \centering\vspace{-10pt}
\includegraphics[width=0.89\linewidth]{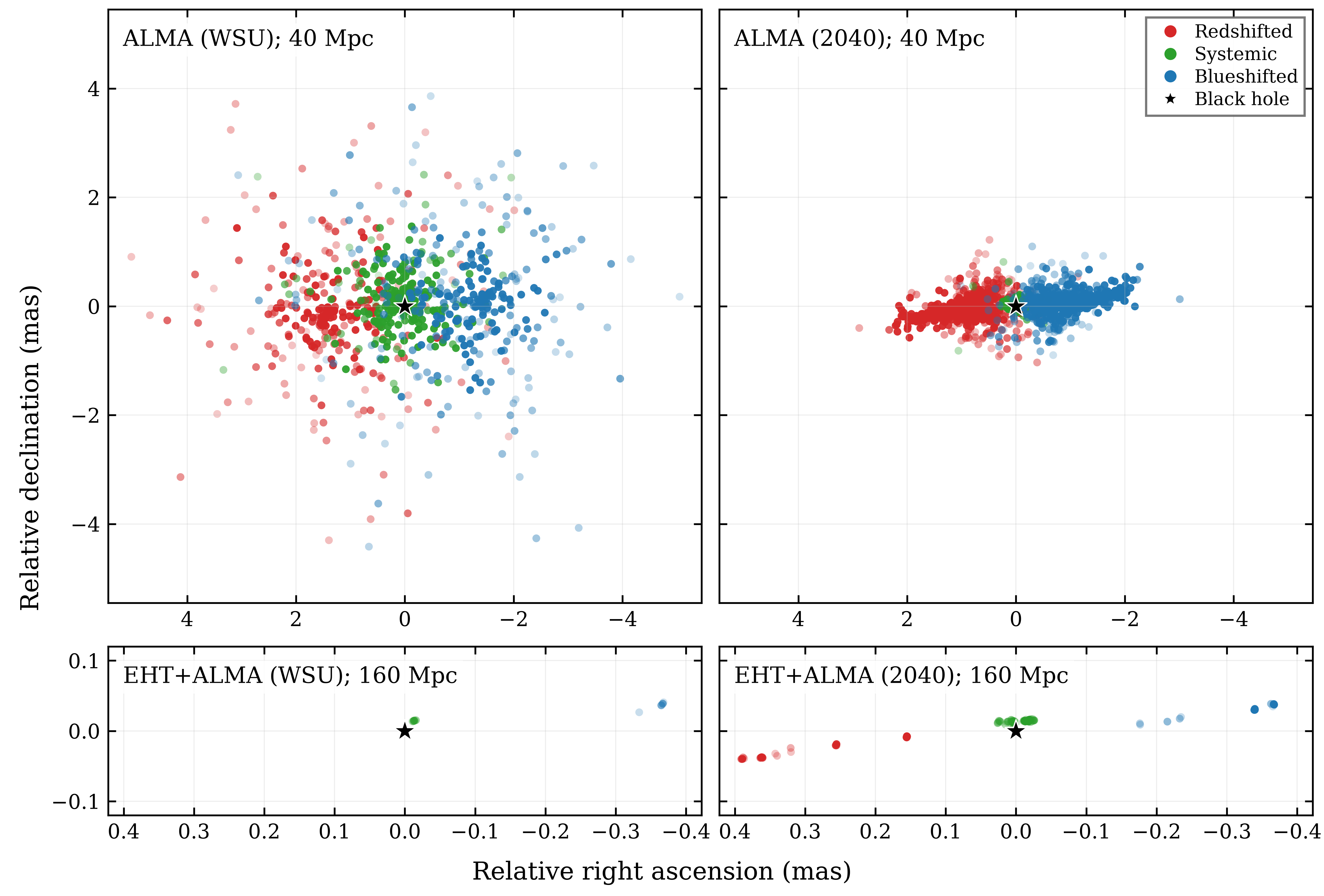}
      \caption{Simulated emission of a NGC~5765b-like megamaser disk at 321\,GHz. Top: Disk at 40\,Mpc comparing current ALMA (left) and ALMA2040 (right). Bottom: the same disk at 160~Mpc, comparing current EHT+ALMA (left) and EHT+ALMA2040+FTP (right).} 
    \label{fig:2} 
\end{figure}

To simulate these maser systems, we generate ground-truth megamaser disk configurations using a standard warped-disk model \citep[e.g.,][]{Herrnstein_1999,Humphreys_2013,Pesce_2020a}, drawing individual redshifted, blueshifted, and systemic maser spot populations from a power-law flux distribution and placing them at radii separations of ${\approx}0.2$\,pc around a $10^7$\,M$_{\odot}$ BH. Each spot is assigned a position and line-of-sight velocity determined by its location inside the disk and the disk's orientation on the sky, for which we adopted an inclination angle of $85^{\circ}$ and a position angle of $90^{\circ}$, with a modest warp in both dimensions. We set the overall luminosity such that the peak flux density in the spectrum is ${\approx}0.2$\,Jy at a distance of 100\,Mpc, similar to the 22\,GHz maser system in NGC~5765b \citep{Gao_2016}.

We conducted synthetic observations of the 321\,GHz maser line for both ALMA2040-only and ALMA2040+EHT VLBI configurations using the {\sc ngehtsim} package \citep{Pesce_2024}. The 321\,GHz transition is particularly favorable: while the 183 and 325\,GHz lines could reach higher flux densities, observations at these frequencies are more strongly affected by atmospheric opacity; the 183\,GHz line is not easily accessible to a large fraction of existing EHT stations. We assume a current ALMA sensitivity (i.e., $1\sigma$ RMS noise level) in a one-hour integration of 4~mJy\,beam$^{-1}$ per 0.5\,km\,s$^{-1}$ channel at 321\,GHz, and we scale the noise level with source redshift. We assume that ALMA2040 has $10\times$ the sensitivity and $3\times$ the angular resolution of current ALMA. For each spectral channel, we define the intrinsic on-sky position as the flux-weighted centroid of all contributing maser spots, to which we then add noise perturbations with $\sigma_{\text{pos}} = 0.6 \left( \frac{\text{FWHM}_{\text{beam}}}{\text{SNR}} \right)$.

ALMA2040 simulations show a significant improvement over ALMA-WSU in both the detection and spatial mapping of megamaser systems. The combination of increased sensitivity and angular resolution enables recovery of the disk spatial and kinematic features, enabling detailed disk modeling and opening the possibility of deriving BH masses and geometric distances directly from sub-mm masers. In this regime, ALMA2040 moves the field from detecting individual sub-mm maser lines to using them as probes of the accretion disk dynamics and physical conditions. For an intrinsically luminous system like NGC~5765b, the simulations show that ALMA2040 could detect the 321\,GHz transition out to $z\approx0.3$, where the line is redshifted into ALMA Band~6. Fig.~\ref{fig:2} shows a simulated disk at 40\,Mpc. However, spatially resolved information is substantially more demanding, and ALMA-only mapping becomes more challenging beyond $z\approx0.05$.

Synthetic ALMA2040+EHT observations show that, for a Circinus-like 321\,GHz maser, detections on the longest baselines are restricted to the nearby Universe even with ALMA2040 and an expanded future EHT array \citep{Doeleman2023, Namene2024IAUGA}. A source beyond $\approx15$\,Mpc may not be recovered on long baselines. This limitation is primarily driven by baseline sensitivity and atmospheric coherence, but may be alleviated only by improved phased-array sensitivity and FPT \citep{Issaoun2025}, i.e. using simultaneous lower-frequency observations. Figure~\ref{fig:2} (bottom right) shows the remarkable ability of ALMA2040 to recover the VLBI disk out to 160\,Mpc, enabled by the improved phase coherence through FPT. The effectiveness of this approach will depend on the multi-frequency capabilities of ALMA2040 and the wider VLBI array.

In an optimistic scenario, combining an intrinsically luminous disk maser similar to NGC~5765b with FPT from bright compact continuum emission at $\approx86$\,GHz, the simulations extend the 321\,GHz VLBI detection horizon up to $\sim300$\,Mpc ($z\sim0.08-0.09$). This gain requires compact 86\,GHz continuum emission at the level of approximately 10--100\,mJy.  It is to be noted that, because of the very small sub-mm VLBI beam, even relatively low-SNR detections of compact maser spots can yield well-constrained positions, allowing informative disk mapping close to the detection limit.


\section{Synergies}

\subsection{LSST/Rubin Observatory}

The Vera C. Rubin Observatory recently began its 10-year Legacy Survey of Space and Time (LSST), which will probe the dynamic Universe by continuously monitoring 18000 deg$^2$ of sky \citep{Ivezic:2019}. LSST will enable constraints on $H_0$ to high precision from standard candles by increasing the number of detected Type 1a supernovae by a factor of 100 out to $z\approx1$ \citep{Jones:2019}, from geometric time delays from hundreds of strong lenses, potentially out to $z\approx5$ \citep{Liao:2019}, and from strongly lensed Type 1a supernovae \citep{Birrer:2022}, among others. These methods require assumptions, both empirical relations and model assumptions (e.g. \citealt{Raney:2020}). Geometric distance measurements from maser discs from ALMA2040 would complement these studies by enabling calibration of standard candles and strong lenses to correct systematic errors. 

In addition, the Rubin Observatory and other current and future multi-wavelength facilities, e.g. JWST, Lynx, LISA, AXIS, will be sensitive to IMBH candidates (e.g. \citealt{Saini:2022}). ALMA2040 observations could provide accurate masses for newly discovered candidates as well as discover additional ones.

\subsection{SKA, ngVLA, and ngEHT}

The forthcoming Square Kilometre Array (SKA) and proposed next-generation Very Large Array (ngVLA) will both provide unparalleled resolution and sensitivity at lower wavelengths \citep{Bonaldi:2026,Selina:2018}. They will be able to push forward studies of 22\,GHz megamasers in the local Universe (ngVLA; \citealt{Braatz:2019}) and redshifted megamasers (SKA; \citealt{Rygl:2026}), complementing ALMA2040 studies of high-frequency water masers. In addition, a planned next-generation EHT (ngEHT; \citealt{Doeleman:2023}) will introduce additional dishes and simultaneous multi-band observing \citep{Johnson:2023} that will pave the way for ALMA2040-VLBI.

\section{Summary}



Sub-mm/mm masers complement the established 22\,GHz technique by probing different excitation conditions and disk regions, constraining the physical conditions of the accreting gas. The reduced free-free absorption in the sub-mm/mm can reveal the far side of the disk, and the order-of-magnitude smaller mm-VLBI beam sharply improves localization of individual maser spots. We propose that ALMA2040 high-resolution, high-sensitivity imaging of nearby disks would yield the most precise BH masses and independent geometric distances, while increased sensitivity would detect sources out to high redshifts. ALMA2040-VLBI of the most local systems would probe fundamentally new gas regions and finer scales than possible at 22\,GHz, resolving compact disks around low-mass BHs. Mapping sub-mm masers across the full disk, including the far side, would constrain properties such as warps, connecting the sub-pc disk to the larger-scale accretion flow and inward to the BH. Water megamaser observations with ALMA2040 will reveal the magnetic structure of AGN accretion disks, discover IMBHs and constrain theories of BH-galaxy coevolution, and provide a precise and accurate measurement of the Hubble constant.


\subsection*{Acknowledgements}
This work was developed as part of the broader ALMA2040 community effort. We acknowledge the Lorentz Center and NOVA for supporting the community workshops and coordination activities that helped initiate and shape the ALMA2040 White Paper process. Support for DWP was provided by the NSF through grants AST-1935980, AST-2034306, and AST-2535855; and by the Gordon and Betty Moore Foundation through grants GBMF5278 and GBMF10423.  This work was supported in part by the Black Hole Initiative at Harvard University, which is funded by the John Templeton Foundation (grants 60477, 61479, and 62286) and the Gordon and Betty Moore Foundation (grant GBMF8273). \\ \bigskip

\raggedright
\bibliographystyle{apj_mod}
{\bibliography{bibliography}}
\end{document}